\documentstyle[psfig]{article}

\begin{document}
\title{
HIGH $p_T$ PHENOMENA IN HEAVY ION COLLISIONS \\
AT $\sqrt{s}=20$ and $200$  AGeV
}

\author{
Miklos Gyulassy$^1$\footnote{Talk given at the 8th
Int. Workshop on Multiparticle Production, June 14-21, 1998,
M\'atrah\'aza, Hungary}
\  and Peter Levai$^{1,2}$\\[2ex]
$^1$Physics Department, Columbia University, New York, NY 10027\\[1ex]
$^2$KFKI Research Institute for Particle and Nuclear Physics, \\
P.O. Box 49, Budapest, 1525, Hungary
}

\date{9 September 1998}
\maketitle

\begin{abstract}
Jet quenching as observed via high $p_T$ hadron production is
studied in the HIJING1.35 model.
We test the model on recent WA98 data on 160 AGeV $Pb+Pb\rightarrow
\pi^0$ up to $4$ GeV/c.
The results  depend sensitively
on the model of the Cronin effect. At (RHIC) collider
energies ($\sqrt{s}>200$ AGeV), on the other hand, semi-hard hadron
production becomes insensitive to the above model
and is thus  expected to be a cleaner probe of jet
quenching signature associated with non-Abelian radiative
energy loss.
\end{abstract}


WA98 $Pb+Pb\rightarrow
\pi^0$ data~\cite{wa98} in the $ 2 \ GeV < p_\perp < 4\  GeV$ range
were analyzed  recently by Wang~\cite{wang98} via a parton model.
He showed that the
 parton model~\cite{owens} is capable  
of reproducing the observed $\pi^0$ invariant inclusive
cross sections  simultaneously in
$p+p$, $S+S$~\cite{wa80}, and central $Pb+Pb$~\cite{wa98}
in the CERN/SPS energy range, $\sqrt{s} \sim 20$ AGeV. 
 That analysis is interesting because it seems to imply
the  absence of quenching of high transverse momenta
hadrons that should arise  if partons lose energy
in dense matter~\cite{MGMP,MGXW92}. 
The $S+S$ and $Pb+Pb$ data clearly reveal, on the other hand,
the expected  Cronin enhancement~\cite{cronin} of moderate
$p_\perp$ pions, as seen first in $p+A$. The $pp$ data
confirm the need to supplement the LLA with non-perturbative intrinsic
$k_\perp$~\cite{feynman,field}. 
We address this problem here, following Ref.~\cite{gyullev},
using the HIJING event generator~\cite{hijing,wang-rep}.

\begin{figure}[t]
\vspace{-1.0in}
  \centerline{\psfig{figure=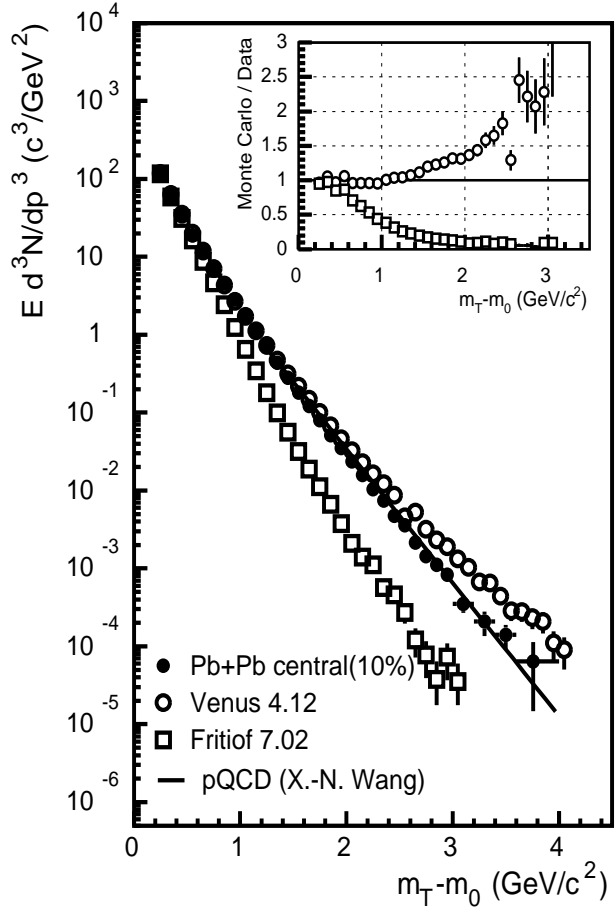,width=3.5in,height=5.0in}}
\vspace{-0.05in}
  \caption{WA98 results~$^1$ on  central (10\%)
  $Pb+Pb\rightarrow \pi^0$ at 158 $A$GeV. Data (solid dots)
  are compared FRITIOF7.02 (open squares)~$^{14}$,
  VENUS4.12 (open circles)~$^{13}$,
  and parton model~$^2$ calculations.
  Inset shows the ratios of the
  results of the Monte Carlo codes to data.
  Figure is from WA98 Coll., Ref.~$^1$.}
  \protect\label{fig:spectra}
\end{figure}

Figure~\ref{fig:spectra} displays the prediction 
of VENUS~\cite{Werner} and FRITIOF~\cite{frit93}
models also for the Pb+Pb collision. 
It can be clearly seen that at high $p_T$ VENUS overpredict
the experimental data of WA98 by a factor of 2.5 and
FRITIOF underpredict it by order of magnitude.
A similar discrepancy was reported
for  S+Au collisions~\cite{wa80}.
The solid curve in Figure~\ref{fig:spectra} 
is the parton model result~\cite{wang98}.

\begin{figure}[bth]
\vspace{-0.5in}
\centerline{\psfig{figure=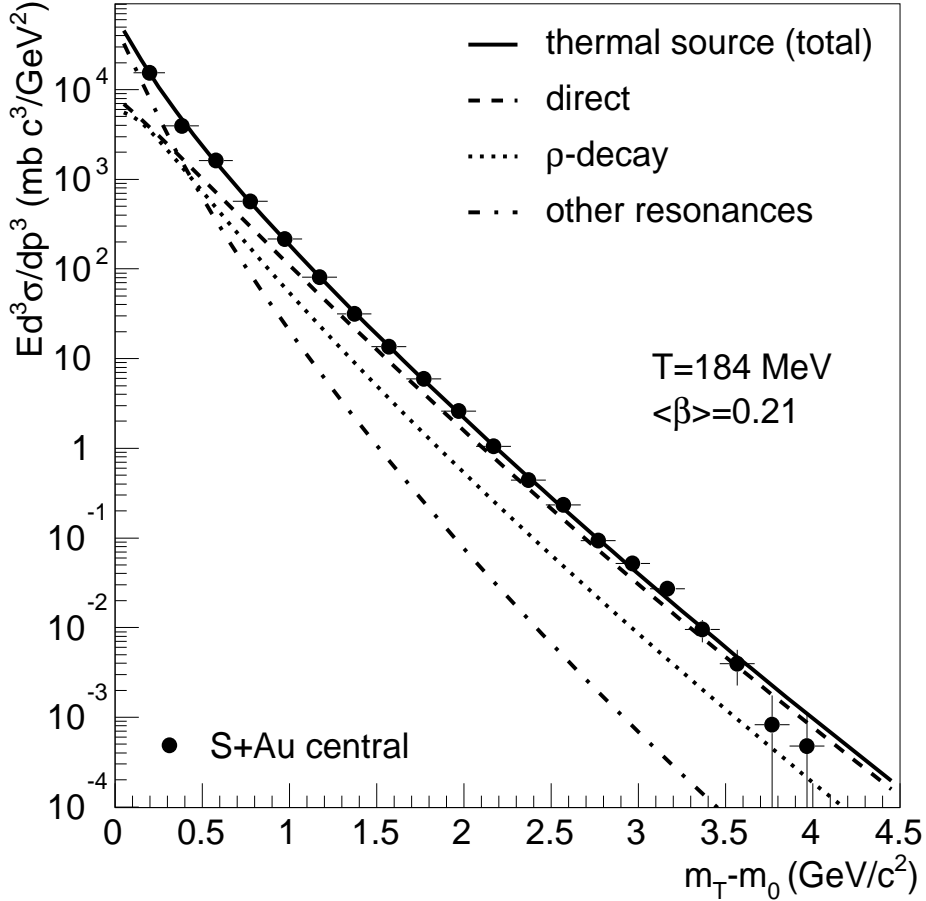,width=5in,height=5.0in}}
\vspace{-0.05in}
\caption{  Boosted Fireball model~$^{15}$ 
fit to $S+Au\rightarrow
\pi^0$ reaction.
  Figure is from WA80 Coll., Ref.~$^4$.}
\protect\label{fig:thermosau}
\end{figure}

We note that the spectra can also be fit well with a simple Boosted Fireball
model~\cite{wied} (see Figure~\ref{fig:thermosau} for S+Au).
However such fits are far less informative than the parton model model
one since the observed $A^{4/3}$ scaling of the high $p_T$ tails
is predicted and in accord with the parton model,
while that dependence is not predicted in the fireball model.
The WA98 data~\cite{wa98}
can also be reproduced by more detailed scaling  hydrodynamic calculations
with suitable initial and final freeze-out criteria~\cite{dumitru}.
Again the real power of pQCD is its prediction without parameters of the $A$
dependence. 

In central heavy ion  reactions at the SPS, conservative estimates
of the initial energy density suggest  that
$\epsilon(1 \;{\rm fm/c}) \approx 1\sim 5$ 
GeV/fm$^3$, are reached. 
The gluon radiative energy loss of partons
in such dense matter is expected to
exceed $dE/dx\sim 1$ GeV/fm~\cite{MGMP}. Recent theoretical
analysis~\cite{BDMPS} predicts in fact a much larger non-linear energy loss 
$\Delta E \sim (\Delta x)^2
$ GeV/fm$^2$  if a parton traverses a quark-gluon plasma
of thickness $\Delta x$.
On the other hand, the WA98 data seem to rule
out $dE/dx>0.1$ GeV/fm, as  Wang showed in 
Ref.~\cite{wang98}. 
In this work, we consider this problem
using the nuclear collision event generator, HIJING~\cite{hijing,wang-rep}.

Recall that in the  LLA, pQCD
predicts that  the single inclusive hadron
cross section is given by Refs.~\cite{owens,field} 
\begin{eqnarray}
  \frac{d\sigma^{AB\rightarrow hX}}{dyd^2p_T}&=&K\sum_{abcd}
  \int d^2\vec{\kappa}_{a} d^2\vec{\kappa}_{b} 
g(\vec{\kappa_a}) g(\vec{\kappa_b}) 
  \nonumber \\ & & \int dx_a dx_b  f_{a/A}(x_a,Q^2)f_{b/B}(x_b,Q^2)
   \nonumber \\ & & 
  \frac{d\sigma}{d\hat{t}}(ab\rightarrow cd) 
  \frac{D^0_{h/c}(z_c,Q^2)}{\pi z_c}
\; \; \; .\label{parton}\end{eqnarray}
This formula convolutes the elementary pQCD parton-parton  cross sections,
$d\sigma(ab\rightarrow cd)$,
 with non-perturbative lowest order fits
of the parton structure functions, $f_{a/A}$, 
and the parton fragmentation functions, $D_{h/c}$,
to $ep$ and $e^+e^-$ data. 
Here ${\vec \kappa}_a, {\vec \kappa}_b$ are the intrinsic transverse
momenta of the colliding partons.
The model includes a $K\approx 2$ factor to simulate
next-to leading order  corrections~\cite{xwke}
at a hard scale $Q\sim p_T/z_c$. The scale dependence
of the structure and fragmentation functions account for the
multiple soft collinear radiative effects. 
However, below energies $\sqrt{s}< 100$ GeV, it is well
known~\cite{feynman} that LLA significantly underpredicts
the $p_\perp<10$ GeV cross section, and additional non-perturbative
effects must be introduced to bring LLA into agreement with data.
Unfortunately, as emphasized in Ref.~\cite{owens}, 
the results are then quite model
dependent  below collider (RHIC) energies.
As we show below, the good news is that this $p_T$ broadening
 dependence
is reduced significantly at collider (RHIC) energies.  

In spite of the inherent ambiguity of the parton model  analysis
at SPS energies, a successful phenomenological
approaches to this problem has been developed
via  the  introduction~\cite{field} of intrinsic
transverse momenta of the colliding partons as in (\ref{parton}).
Originally, a 
Gaussian form for that distribution 
with $\langle k_\perp^2\rangle\sim 1$ GeV$^2$ was proposed in
Ref.~\cite{field}.
However, in order to reproduce $pp$ data more accurately and to take into 
account the Cronin effect, 
the Gaussian ansatz was generalized in Ref.~\cite{wang98}
to include $Q^2$ and $A$ dependence. With
 $g(\vec{\kappa})\rightarrow
g_a(\vec{\kappa},Q^2,A)$, 
 excellent fits~\cite{wang98}
to the WA98 data could be obtained assuming a factorized Gaussian
distribution with
\begin{equation}
\langle \kappa^2(b)\rangle_A= \left(1  + 0.2\;
Q^2\alpha_s(Q^2)+ \frac{0.23\;\sigma_{pp}t_A(b)\;\ln^2Q}{1+\ln Q}
\right)  (\rm{GeV/c})^2
\label{kick}\end{equation}
where $Q^2$ is measured here in (GeV/c)$^2$. In (\ref{kick})
$\sigma_{pp}t_A(b)$ is the average number of inelastic scatterings
a nucleon suffers traversing nucleus $A$ at impact parameter $b$.
The nuclear thickness function, $t_A(b)$, is 
 normalized as usual to $\int d^2b t_A(b)=A$. 
Eq.(\ref{kick}) is the main source of the model dependence
in Ref.~\cite{wang98}.

The HIJING1.35 Monte Carlo model~\cite{hijing,wang-rep}
 incorporates pQCD jet production together with initial and final state
radiation according to the PYTHIA algorithm~\cite{pythia}.
In addition, it incorporates  a variant
of the soft string phenomenology similar
to the Lund/FRITIOF~\cite{fritiof}
 and DPM~\cite{dpm94} models to simulate beam jet fragmentation
and jet hadronization physics. Low transverse momenta inelastic processes
are  of course highly model dependent, and the 
parameters must be fit to 
$pp$ and $AB$ data~\cite{hijing,wang-rep}. 
It is of interest to apply
HIJING to  the present study
because it incorporates in addition to the above
soft and hard dynamics,  a model of soft (Cronin)
multiple initial state  collision effects 
as well as a simple jet quenching scheme. With these features, we are able
to study how competing aspects of the reaction mechanism
influence hadronic observables and explore the magnitude of
 theoretical uncertainties.

\begin{figure}[t]
\centerline{
\psfig{figure=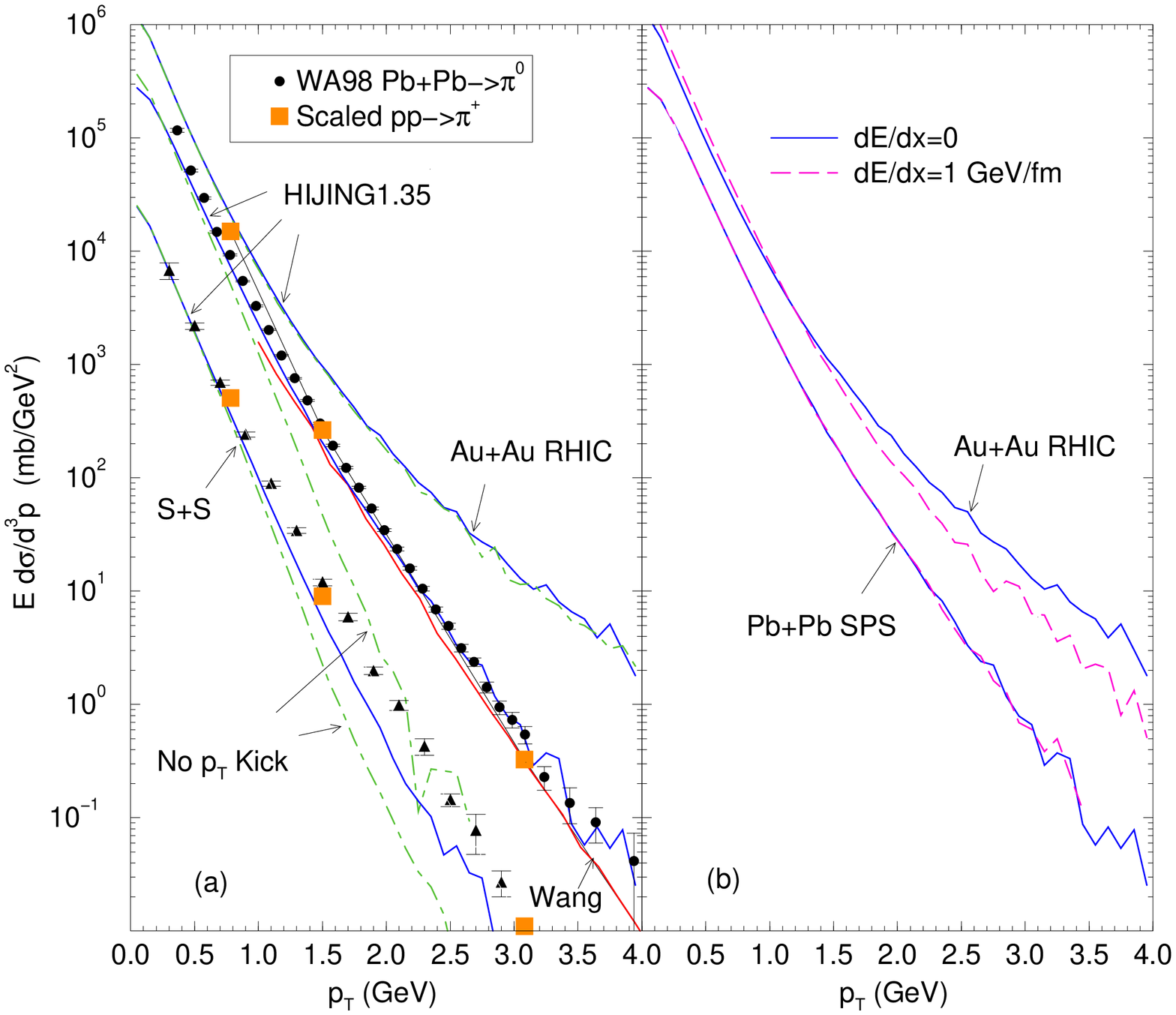,width=4.6in,height=4.5in,angle=-90.}}
\caption{ Invariant \protect{$A+A\rightarrow \pi^0$} cross section
for central collision at SPS and  RHIC energies are compared.
a) The WA80 $S+S$ data~$^4$ (triangles) and 
the WA98 $Pb+Pb$ data~$^1$ 
(dots) are compared to HIJING1.35~$^{11}$
with soft $p_\perp$ kicks (full lines)
and without $p_T$ kicks (dot-dashed curves). The later  scale
with the wounded projectile number
times \protect{$\sigma_{AA}$} times
the invariant distribution calculated for \protect{$pp$}.
The parton model curve is from Ref.~$^2$
is labeled by 'Wang'. 
The filled squares show \protect{$pp\rightarrow \pi^+$} data
scaled by the (Glauber) number of binary collisions 
times \protect{$\sigma_{AA}$} for both $SS$ and $PbPb$.
 b) Jet quenching, predicted  at RHIC energies~$^6$,
is not significant at SPS energies in the HIJING model. 
}
\label{fig3}
\end{figure}

In the HIJING model, excited baryon string
 are assumed to
pick up random transfer momentum kicks in each inelastic
scattering according to the following
distribution
\begin{equation}
g(\vec{\kappa})\propto \left\{(\kappa^2+p_1^2)(\kappa^2+p_2^2)(1
+ e^{(\kappa-p_2)/p_3})\right\}^{-1}\;\;,\label{hjpt}
\end{equation}
where $p_1=0.1,\  p_2=1.4,\  p_3=0.4$ GeV/c 
were chosen to fit low energy ($p_\perp<1$ GeV/c)
multiparticle production data 
in Refs.~\cite{hijing,wang-rep}.
A flag, IHPR2(5)(=1 or 0),  
makes it possible to compute spectra with and without 
this effect as shown in part (a) of Figure~\ref{fig3}.
The present study is the first test of this model up to $4$ GeV/c
in the SPS energy range.

Jet quenching is modeled via gluon splitting according to the
number of mean free paths, $\lambda=1$ fm, traversed by a gluon
through the cylindrical nuclear reaction volume. 
In each partonic  inelastic interaction 
a gluon of energy $\Delta E= \Delta x \; dE/dx$ is assumed to be split off
the parent jet and incorporated
as a kink in another baryonic string~\cite{hijing}.
The (constant) energy loss per unit length is 
an input parameter (HIPR1(14) in HIJING
~\cite{hijing}) and
 can switched on and off via IHPR2(4) (=1 or 0) to test the 
sensitivity of spectra to jet quenching as shown in Figure~\ref{fig3}.

Figure~\ref{fig3} compares   the predictions
of HIJING1.35~\cite{hijing,wang-rep} without jet quenching
($dE/dx=0$) for the invariant
$\pi^{0}$ cross section in central nuclear collisions
at SPS and RHIC energies. 
The cross section for  central $A+A$ collisions
are computed integrating over the impact
parameters up to $b_{\rm max}$ chosen to reproduce
experimental trigger cross sections. 
For WA98 $Pb+Pb$ and RHIC $Au+Au$ we took
$b_{\rm max}=4.5$ fm, while for the WA80 $S+S$ we took
$b_{\rm max}=3.4$ fm. The multiple collision eikonal geometry in HIJING
is based on  standard Wood-Saxon nuclear densities.

The  parton model fit to the WA98 data are
labeled by 'Wang' from Ref.~\cite{wang98}. (The normalization of both
the WA98 data and Wang's latest calculations (not shown)
have increased $\sim 20-40\%$  relative to Ref.~\cite{wang98}.)
We note that the HIJING1.35 calculation for this interaction
given by the solid jagged line also reproduces remarkably well
the $\pi^0$ invariant cross sections without jet quenching. 
However, for the lighter $S+S$ reaction,
HIJING, underestimates the $p_\perp >1$ GeV/c tail significantly. 
This error is traced to the failure of the model to
reproduce the $pp$ high $p_\perp$ data at these energies, in contrast
to its successful account of higher energy data~\cite{hijing}.
This can be seen by comparing the filled squares to the dot-dashed
curves as explained below.
Therefore, we find that the agreement with the WA98 data
is accidental and the observed $A$ scaling
of the high $p_\perp$ region at SPS energies is not reproduced.

Figure~\ref{fig3}a shows that the soft transverse momentum kick model is
the source for the 
agreement of HIJING with the $Pb+Pb$ data.  The dot-dashed curves show
what happens if the soft $p_\perp$ kicks modeled with eq.(\ref{hjpt})
is turned off. The very strong decrease of the pion yield
in the $Pb+Pb$ case and the somewhat smaller but still large decrease in the
$S+S$ case shows clearly the important
role multiple transverse kicks   at these
 energies. In fact 
both $S+S$ and $Pb+Pb$ dot-dashed curves are found
to coincide with the calculated  $pp\rightarrow \pi$
differential cross section scaled by 
the wounded nucleon factor $W_A \sigma_{AA}/\sigma_{NN}$,
where $W_A=21 , 172$ is the average number of wounded projectile
nucleons and $\sigma_{AA}=32,363,636$ mb for $A=1,32,207$.

On the other hand, the data follow closely the shape of the 
measured $pp\rightarrow \pi^+$ data taken from \cite{wang98}
 and scaled to $AA$ by multiplying by  the
Glauber binary collision number factor, $T_{AA}\sigma_{AA}$.
From HIJING the average number of binary
collisions in these systems is $45$, $751$ resp.
The fact that the WA98 data scale with the above Glauber factor
within a factor of three, suggests that the additional
$p_\perp$ broadening due to initial state collisions is relatively small.
The very large $A$ dependence of the HIJING $p_\perp$ tail
at SPS energies is due to the $A^{1/3}$ times convolution of the distribution
(\ref{hjpt}). This problem is avoided in the parton model
calculation~\cite{wang98} using (\ref{kick}) through the separation of 
larger intrinsic momentum effects and smaller $A^{1/3}$
dependent contributions.
 
We conclude that the missing intrinsic transverse
momentum component of HIJING precludes an accurate simultaneous
reproduction of of $pp,SS,PbPb$ data at SPS energies.
However, unlike the parton model where no global conservation
laws have to be enforced, it is not clear how to incorporate intrinsic
momenta in a global event generator like HIJING
without destroying the satisfactory
reproduction
of {\em low} $p_\perp<1$ GeV/c data. We do not attempt 
to solve  this problem here.

Our main point is that this problem goes away
fortunately at higher (RHIC) collider energies ($\sqrt{s}=200$ 
AGeV).
As seen in Figure~\ref{fig3}a the effect of multiple soft interactions is 
very much reduced at that energy.
This is due to the well known effect that as the beam energy increases,
the $p_\perp$ spectra become harder and additional $p_\perp$
smearing  from initial state effects becomes relatively less 
important. It is the extreme steepness of the cross sections at SPS energies
that amplifies so greatly the sensitivity of the moderate $p_\perp$
yields to this aspect of multiple collision dynamics. 

In Figure~\ref{fig3}b we consider next the sensitivity of 
the pion yields to the sought after
parton energy loss dynamics.
The striking difference between Figure~\ref{fig3}a 
and Figure~\ref{fig3}b is that
in Figure~\ref{fig3}b the SPS yield is not sensitive 
to the energy loss model
in HIJING, 
while at RHIC energies the suppression of semi-hard hadrons
is seen to be sensitive to  jet quenching as predicted in Ref.~\cite{MGXW92}.
This seems counter intuitive at first because
the increasing steepness with decreasing beam energy is naively
expected to result in greater quenching for a fixed jet energy loss.
However, in this model  the observed $p_\perp$ range is 
dominated
by multiple soft collisions.
The moderate $p_\perp<5$ GeV
quarks, which fragment into the observed pions,
are not produced
in rare pQCD semi-hard interactions but are gently nudged several times
 into
that $p_\perp$ range. Since in HIJING
jet-quenching is restricted
to only those partons that suffer  a semi-hard pQCD interaction with $p_\perp$
at least $p_0=2$ GeV/c (the mini-jet scale), no quenching arises at this
energy.

In conclusion, we found that HIJING1.35 fits the WA98 data
with or without jet quenching. 
However, this model fails to account for the weak $A$ dependence of the
data that scale approximately with the Glauber $T_{AA}$ factor,
and therefore the fit must be viewed as accidental.
The hydrodynamic model fits~\cite{dumitru} of WA98 data~\cite{wa98}
unfortunately make no prediction for the A dependence. 
On the other
hand, another Monte Carlo parton cascade model~\cite{VNI}
overpredicted the $\pi^0$ cross section at $4$ GeV/c by a factor
of 10.  Those results and ours  emphasize 
the strong model dependence of the SPS spectra 
due to non-perturbative aspects of the problem. Those
aspects, while phenomenologically interesting, make it difficult to 
identify perturbative QCD phenomena
and search for  jet quenching.

The main point of this work
is to emphasize the good news that at collider energies there is  much less
sensitivity to the above uncertain element of the reaction mechanism.
At $\sqrt{s}> 100$ AGeV the expected~\cite{MGXW92}
jet quenching signature should be readily observable in the $p_\perp\sim 10$
GeV range. Such experiments will commence in 1999 at RHIC 
and should  provide
important tests of the theory of non-Abelian multiple collisions and energy
loss~\cite{MGMP,BDMPS}. 

\section*{Acknowledgments}
We would like to thank T. Cs\"org\H o for the kind hospitality  
at M\'atrah\'aza.
We also thank X.N. Wang, K. Eskola and D. Rischke
 for the extensive discussions, 
and T. Peitzmann for making the WA98 preliminary data available.
This work was supported by the Director, Office of Energy Research,
Division of Nuclear Physics of the Office of High Energy and
Nuclear Physics of the U.S. Department of Energy under Contract No.
DE-FG02-93ER40764 and partly by US/Hungarian Science and 
Technology Joint Fund No.652/1998. and the OTKA Grant
No.T025579.

\end{document}